\documentclass[aps,prd,floatfix,nofootinbib]{revtex4} % preprint, twocolumn, showpacs,groupedaddress,

\usepackage{graphics}
\usepackage{graphicx}

\usepackage{latexsym}
\usepackage[cp866]{inputenc}
\usepackage[english]{babel}

\input epsf

\begin{document}

\title{\boldmath Triangle singularities in the $T^+_{cc}\to D^{
*+}D^0\to\pi^+D^0D^0$ decay width}
\author{N.N. Achasov \footnote{achasov@math.nsc.ru}
and G.N. Shestakov \footnote{shestako@math.nsc.ru}
}\affiliation{Laboratory of  Theoretical Physics, S.L. Sobolev
Institute for Mathematics, 630090, Novosibirsk, Russia}

%\date{}

\begin{abstract}
%----------------------------------------------------------------------------------------------
The values of the masses of the particles involved in the decay of
$T^+_{cc}\to D^{*+}D^0\to\pi^+D^0D^0$ suggest that due to the final
state interactions in the transition vertex $T^+_{cc}\to D^{*+}D^0 $
there may be triangle logarithmic singularities. We discuss their
possible role and show that the tree approximation for calculating
the decay widths $T^+_{cc}\to(D^{*+}D^0+D^{*0}D^+)\to\pi^+D^0D^0$,
$\pi^0D^0 D^+$, $\gamma D^0D^+$ is quite sufficient at the current
level of measurement accuracy.
%----------------------------------------------------------------------------------------------
\end{abstract}

\maketitle

\section{Introduction}

At the end of July 2021, the LHCb Collaboration announced the
discovery of the doubly charmed tetraquark $T^+_{cc}$ \cite{Muh21,
Pol21} and then published detailed measurement results along with
their theoretical processing and interpretation \cite{Aa21a,Aa21b}.
Over the next days, weeks and months, a very interesting discussion
is going on in the literature about the possible internal structure
of the $T^+_{cc}$ state, about the mechanisms of its production and
decay through intermediate states $D^{*+}D^0$ and $D^{*0}D^+$, on
the possible values of its total decay width and partial decay
widths into coupled channels $\pi^+D^0D^0 $, $\pi^0 D^0 D^+$, and
$\gamma D^0D^+$, about its line shape and the shapes of the
two-particle mass spectra $DD$ and $\pi D$, as well as about the
possible existence of other similar states. A detailed discussion of
all these issues can be found in \cite{Li21,Me21,FLO21,Al21,YV22,
Guo22,Bra22,HL22,Lin22,Ch21,Mikh22,Qin21,Meng22}; see also the
references cited therein.

In Sec. II of this article, we discuss the possible role of triangle
singularities in the width of the decay $T^+_{cc}[3.875;I(J^P)=
0(1^+)]\to D^{*+}(J^P= 1^-)D^0(J^P=0^-)\to \pi^+D^0D^0$. In so
doing, we proceed within the framework of a scalar model, i.e., we
treat all particles in this decay as spinless and scalar. Such a
simplification, however, seems quite reasonable. First, the decay of
$T^+_{cc}\to D^{*+}D^0$ occurs in the near-threshold region and
therefore is mainly $S$ wave. Second, the ratio of the
$D^{*+}\to(D\pi)^+$ decay width (it is $\approx83$ keV \cite{PDG20})
to the distance to the $(D\pi)^+$ threshold is $\approx1 /70$ and,
consequently, the change of this width on the energy interval of the
order of itself is small (i.e., in the $D^{*+}(2010)$ region, it is
almost constant). The formulas [see below Eqs. (\ref{Eq3}) and
(\ref{Eq4})], which we use to estimate the possible role of
interactions of $D^*D$ pairs in the final state, are in fact
expansions of the Omn\`{e}s functions (solutions) for form factors
\cite{Om58} in case of weak coupling (i.e., smallness of $D^*D$
scattering at low energies). Discussions of a number of dynamic
approximations of the Omn\`{e}s functions can be found, for example,
in Refs. \cite{Bar65,RT81}. The performed analysis allows us to
conclude that the tree approximation used in Refs. \cite{Aa21a,
Aa21b} for calculating the decay widths $T^+_{cc}\to(D^{*+}D^0+
D^{*0}D^+)\to \pi^+ D^0D^0,\pi^0D^0 D^+,\gamma D^0D^+$ is quite
sufficient at the current level of measurement accuracy.

The LHCb Collaboration results \cite{Aa21a, Aa21b} obtained from the
fit to the $\pi^+D^0D^0$ mass spectrum indicates that the
Breit-Wigner mass of the $T^+_{cc}$ relative to the $D^{*+}D^0$ mass
threshold $\delta m_{BW}=-273\pm61$ keV and its Breit-Wigner width
$\Gamma_{BW}=410\pm 165$ keV (only statistical uncertainties are
indicated here). The measured $\delta m_{BW}$ value corresponds to a
mass of approximately 3875 MeV.

%-----------------------------------------------------------------------------------------------

\section{\boldmath{$T^+_{cc}\to D^{*+}D^0\to\pi^+D^0D^0
$} decay in the scalar model}

\begin{figure} % [!ht] % \vspace*{2mm} % width=12cm
\begin{center}\includegraphics[width=12cm]{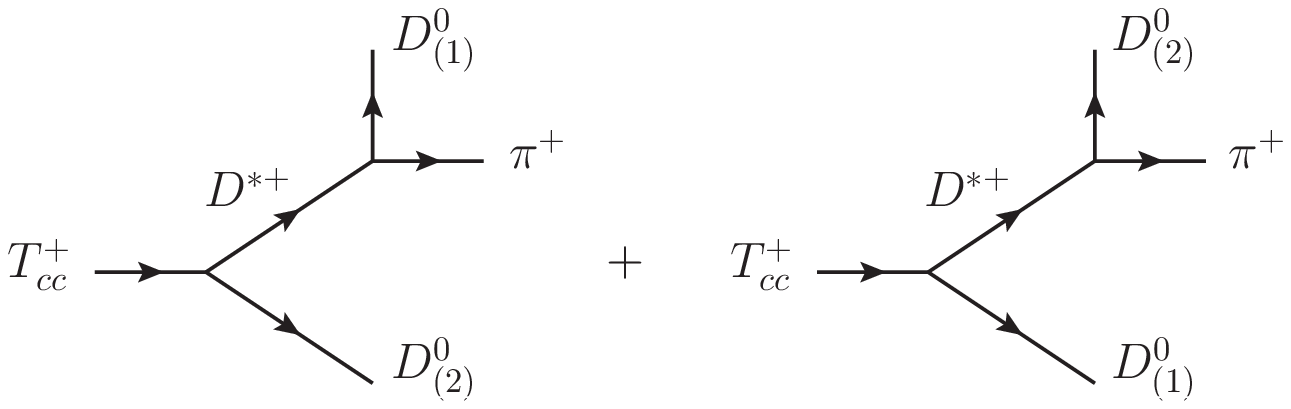}
\caption{\label{Fig1} The tree diagrams for the decay $T^+_{cc}\to
D^{*+}D^0\to\pi^+D^0D^0 $.}
\end{center}\end{figure}

\begin{figure} [!ht] % \vspace*{2mm} % width=9cm
\begin{center}\includegraphics[width=9cm]{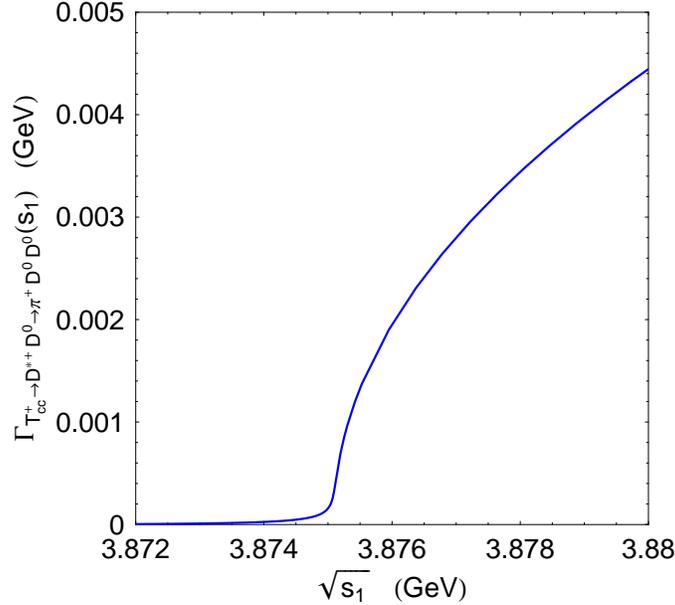}
\caption{\label{Fig2} The $T^+_{cc}\to D^{*+}D^0\to\pi^+D^0D^0$
decay width as a function of $\sqrt{s_1}$ calculated by Eqs.
(\ref{Eq1}) and (\ref{Eq2}).}\end{center}\end{figure}

In the tree approximation, the $T^+_{cc}\to D^{*+}D^0\to\pi^+D^0D^0
$ decay is described by two diagrams shown in Fig. \ref{Fig1} which
differ in the permutation of identical $D^0$ mesons. The
corresponding decay width is given by
\begin{eqnarray}\label{Eq1}
\Gamma_{T^+_{cc}\to D^{*+}D^0\to\pi^+D^0D^0}(s_1)=\frac{
g^2_{T^+_{cc}D^{*+}D^0}}{16\,\pi}\frac{g^2_{D^{*+}\pi^+D^0}}{
16\,\pi}\frac{1}{\,\,2\pi\,s_1^{3/2}}%\nonumber\\ \times
\int\limits_{(m_{D^0}+m_{\pi^+})^2}^{(\sqrt{s_1}-m_{D^0})^2}
ds\int\limits_{t_-(s_1,s)}^{t_+(s_1,s)}dt\
\left|\frac{1}{D(s)}+\frac{1}{D(t)}\right|^2,\qquad\end{eqnarray}
where $s_1$ is the invariant mass squared of the virtual $T^+_{cc} $
state, $s$ and $t$ are the $\pi^+D^0_{(1)}$ and $\pi^+D^0_{(2)} $
invariant mass squared, respectively, and $t_{\pm}(s_1,s)$ denote
the boundaries of the physical region for the variable $t$ for fixed
values of $s_1$ and $s$ \cite{MT21}; $g_{T^+_{cc}D^{*+}D^0}$ and
$g_{D^{*+}\pi^+D^0 }$ are the effective coupling constants. The
inverse propagator of the $D^{*+}$ resonance, $D(s)$, has the form
\begin{eqnarray}\label{Eq2}
D(s)=m^2_{D^{*+}}-s-i\sqrt{s}\Gamma^{tot}_{D^{*+}}(s)=
m^2_{D^{*+}}-s-i\frac{g^2_{D^{*+}\pi^+D^0}}{16\pi}\left[\rho_{
\pi^+D^0}(s)+\frac{1}{2}\,\rho_{\pi^0D^+}(s)\right],
\end{eqnarray}
%--------------------------------------------------------------------------------
%--------------------------------------------------------------------------------
where $\rho_{\pi D}(s)=\sqrt{s^2-2s(m^2_\pi+m^2_D)+ (m^2_\pi-m^2_D
)^2}/s$; factor $1/2$ at $\rho_{\pi^0D^+}(s)$ follows from the
isotopic symmetry. We neglect the contribution of the $D^{*+}\to
D^+\gamma$ decay, which is $\approx1.6\%$. From the Particle Data
Group data \cite{PDG20} about $\Gamma^{tot}_{D^{*+}}(m^2_{D^{*+}})
=83.4$ keV we find $g^2_{D^{*+}\pi^+D^0}/(16\pi)\approx0.00289$
GeV$^2$. We also set $g^2_{T^+_{cc}D^{*+}D^0}/(16\pi)=0.5$ GeV$^2$
for definiteness. A comprehensive discussion of possible values of
this constant which it can be extracted from the fit of the
$T^+_{cc}$ line shape is presented in Ref. \cite{Aa21b}. Let us note
that the relative magnitude of the effect that we consider here does
not depend on the value of the constant $g_{T^+_{cc}D^{*+}D^0}$. The
energy dependent width $\Gamma_{T^+_{cc}\to D^{*+}D^0\to
\pi^+D^0D^0}(s_1)$ calculated in the tree approximation using Eqs.
(\ref{Eq1}) and (\ref{Eq2}) is shown in Fig. \ref{Fig2}. Note that
for large values of $g_{T^+_{cc}D^{*+}D^0}$ the sharp increase of
$\Gamma_{T^+_{cc} \to D^{*+}D^0\to\pi^+D^0D^0}(s_1)$ and
$\Gamma_{T^+_{cc}\to(D^{*+} D^0+D ^{*0}D^+)\to \pi^0D^0D^+}(s_1)$
above the $D^*D$ thresholds and a similar growth of the real part of
the self-energy function under the $D^*D$ thresholds in the
$T^+_{cc}$ propagator lead to a sharp suppression of the right and
left wings of the $T^+_{cc}$ resonance peak. A similar phenomenon
takes place for the four-quark $a_0(980)$ resonance \cite{ADS80}.
%--------------------------------------------------------------------------------
%--------------------------------------------------------------------------------

\begin{figure} %[!ht] % \vspace*{2mm}
\begin{center}\includegraphics[width=14cm]{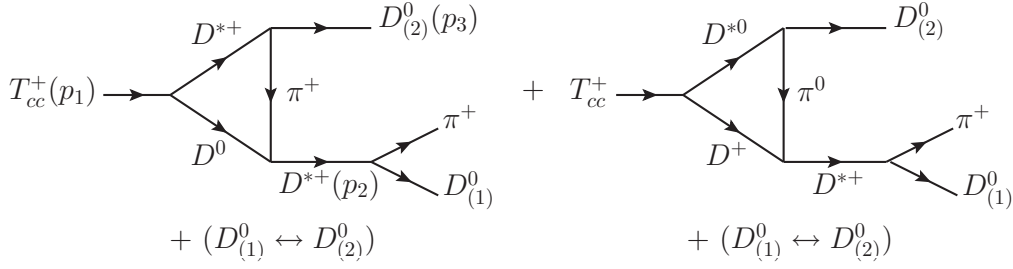}
\caption{\label{Fig3} The simplest diagrams that take into account
final state interactions in the $T^+_{cc}\to D^{*+}
D^0\to\pi^+D^0D^0$ decay.}\end{center}\end{figure}

The values of the masses of the particles involved in the decay of
$T^+_{cc}\to D^{*+}D^0\to\pi^+D^0D^0$ suggest that due to the final
state interactions in the vertex transition $T^+_{cc}\to D^{*+}D^0$
there may be triangle logarithmic singularities. Examples of
corresponding ``dangerous'' diagrams are shown in Fig. \ref{Fig3}.
Take into account their contribution to the decay width
$\Gamma_{T^+_{c c}\to D^{*+}D^0 \to\pi^+D^0D^0}(s_1)$ can be done by
the following substitutions in Eq. (\ref{Eq1}):
\begin{eqnarray}\label{Eq3}
\frac{1}{D(s)}\,\to\,\frac{1}{D(s)}\left[1+\frac{g^2_{D^{*+}\pi^+D^0
}}{16\,\pi}\,\left(F_{D^{*+}D^0}(s_1,s)+\frac{1}{2}\,F_{D^{*0}D^+}
(s_1,s)\right)\right],\\ \label{Eq4} %\nonumber\\
\frac{1}{D(t)}\,\to\,\frac{1}{D(t)}\left[1+\frac{g^2_{D^{*+}\pi^+D^0}
}{16\,\pi}\,\left(F_{D^{*+}D^0}(s_1,t)+\frac{1}{2}\,F_{D^{*0}D^+}
(s_1,t)\right)\right],\end{eqnarray}
%--------------------------------------------------------------------------------
%--------------------------------------------------------------------------------
where $F_{D^{*+}D^0}$ and $F_{D^{*0}D^+}$ are the amplitudes of the
triangle loops included in the diagrams in Fig. \ref{Fig3}; the
factor $1/2$ at $F_{D^{*0}D^+}$ follows from isotopic symmetry. In
our normalization
\begin{equation}\label{Eq5}
F_{D^{*+}D^0}(s_1,s)=\frac{i}{\pi^3}\int\frac{d^4k}{D_1D_2D_3}\,,
\end{equation} where
$D_1=k^2-m^2_{D^{*+}}+i\varepsilon$, $D_2=(p_1-k)^2-m^2_{D^0}
+i\varepsilon$, and $D_3=(k-p_3)^2-m^2_{\pi^+}+i\varepsilon$ are the
inverse propagators of the particles in the loop. Here and in Fig.
\ref{Fig3} $p_1$, $p_2$, $p_3$ denote the 4-momenta of the particles
in the reaction; $p_1= p_2+p_3$, $p_1^2=s_1$, $p_2^2=s$, and
$p_3^2=m^2_{D^0}$. The amplitude $F_{D^{*0}D^+}(s_1, s)$ has a
similar form.

If the values of the variables $\sqrt{s}$ and $\sqrt{s_1}$ are
simultaneously in the intervals
\begin{eqnarray}\label{Eq6}
m_1+m_2<\sqrt{s_1}<\sqrt{m^2_1+m^2_2+m_2m_3+(m_2/m_3)(m^2_1-m^2_{D^0})},
\\ \label{Eq7} %\nonumber\\
m_2+m_3<\sqrt{s}<\sqrt{[(m_1+m_2)(m_1m_2+m^2_3)-m_2m^2_{D^0}]/m_1},\
\ \ \end{eqnarray} where $m_1$, $m_2$, and $m_3$ are the particle
masses in the inverse propagators $D_1$, $D_2$, and $D_3$,
respectively [see Eq. (\ref{Eq5})], then for each value of $s$,
there is a unique value $s_1$ (and vice versa) for which the
imaginary part of the amplitude $F_{D^{*+}D^0}(s_1, s)$ [and
similarly that of $F_{D^{*0}D^+}(s_1,s)$] has a logarithmic
singularity (see, for example, Refs. \cite{FN64,BAGO16,AS19,Guo20}
and references herein). Note that the minimum value of $s_1$ in
(\ref{Eq6}) herewith corresponds to the maximum value of $s$ in
(\ref{Eq7}) (and vice versa). In the amplitude
$F_{D^{*+}D^0}(s_1,s)$ (see the left diagram in Fig. \ref{Fig3})
$m_1=m_{D^{*+}}$, $ m_2=m_{D^0}$, $m_3=m_{\pi^+}$ and for it the
numerical values of the intervals in Eqs. (\ref{Eq6}) and
(\ref{Eq7}) are as follows:
\begin{eqnarray}\label{Eq8}
3.8751\ \mbox{GeV}<\sqrt{s_1}<3.91259\ \mbox{GeV}\quad (0<\sqrt{
s_1}-(m_{D^{*+}}+m_{D^0})<37.49\ \mbox{MeV}),\\ \label{Eq9} 2.00441\
\mbox{GeV}<\sqrt{s}<2.00946\ \mbox{GeV}\quad (0<\sqrt{s}-(m_{D^0}+
m_{\pi^+})<5.05\ \mbox{MeV}).\quad\end{eqnarray} Here in parentheses
are the corresponding intervals in units of MeV into which the
invariant masses $\sqrt{s_1}$ and $\sqrt{s}$ with the subtracted
threshold values should fall. In the amplitude
$F_{D^{*0}D^+}(s_1,s)$ (see the right diagram in Fig. \ref{Fig3})
$m_1=m_{D^{*0}}$, $ m_2=m_{D^+}$, $m_3=m_{\pi^0}$ and for it,
respectively, we have
\begin{eqnarray}\label{Eq10}
3.87651\ \mbox{GeV}<\sqrt{s_1}<3.92318\ \mbox{GeV}\quad (0<\sqrt{
s_1}-(m_{D^{*0}}+m_{D^+})<46.67\ \mbox{MeV}),\\ \label{Eq11}
2.00464\ \mbox{GeV}<\sqrt{s}<2.01073\ \mbox{GeV}\quad (0<\sqrt{s}
-(m_{D^+}+m_{\pi^0})<6.05\ \mbox{MeV}).\quad\end{eqnarray}
%--------------------------------------------------------------------------------
\begin{figure} %[!ht] % \vspace*{2mm}
\begin{center}\includegraphics[width=16cm]{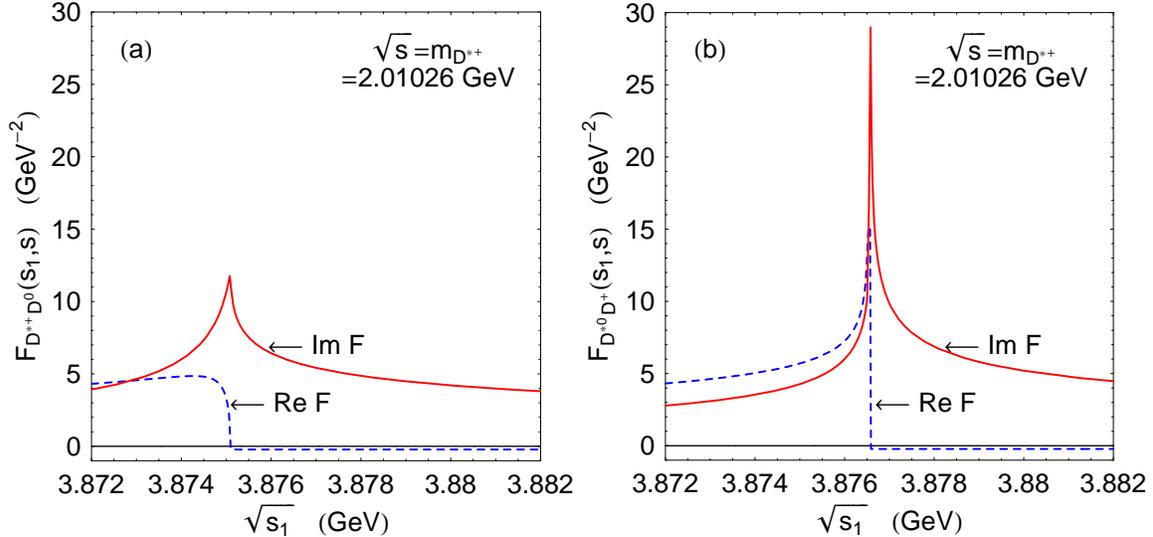}
\caption{\label{Fig4} The solid and dashed curves show,
respectively, imaginary and real parts of the amplitudes (a)
$F_{D^{*+}D^0}(s_1,s)$ and (b) $F_{D^{*0} D^+}(s_1,s)$ for
$\sqrt{s}=m_{D^{*+}}$. In the vicinity of the $D^{*+}$ resonance
(i.e., at $\sqrt{s}\approx m_{D^{*+}}$) the functions $F_{D^{*+}D^0}
(s_1,s)$ and $F_{D^{*0} D ^+}(s_1,s)$ have a similar behavior.}
\end{center}\end{figure}
%--------------------------------------------------------------------------------
Since $m_{D^{*+}}=2.01026\ \mbox{GeV}$ is greater than $(\sqrt{s}
)_{max}=2.00946\ \mbox{GeV}$ in Eq. (\ref{Eq9}), then due to the
contribution of the amplitude $F_{D^{*+}D^0}(s_1,s\approx m^2_
{D^{*+}})$ in the vertex $T^+_{cc} D^{*+}D^0$ no triangle
singularity arises. At the same time, $m_{D^{*+}}=2.01026\
\mbox{GeV}$ is less than $(\sqrt{s})_{max}= 2.01073\ \mbox{GeV}$ in
Eq. (\ref{Eq11}) and therefore in the amplitude $F_{D^{*0}D^+}(s_1,s
\approx m^2_{D^{*+}})$ (and hence in the $T^+_{cc}D^{*+} D^0$
vertex) the triangle singularity occurs at $\sqrt{s_1}\approx
3.87658$ GeV. Figure \ref{Fig4} shows the imaginary and real parts
of the amplitudes $F_{D^{*+}D^0}(s_1,s= m^2_{D^{*+}})$ and
$F_{D^{*0} D^+}(s_1,s=m^2_{D^{*+}})$ as functions of $\sqrt{s_1}$
calculated using the method developed in Refs. \cite{tHV79,Den07}.
As can be seen from this figure, the functions $F_{D^{*+}D^0}(s_1,s=
m^2_{D^{*+}})$ and $F_{D^{*0}D^+}(s_1,s=m^2_{D^{*+}})$ are not small
by themselves (one of them is even singular). However, in Eqs.
(\ref{Eq3}) and (\ref{Eq4}), the contributions from
$F_{D^{*+}D^0}(s_1,s)$ and $F_{D^{*0}D ^+}(s_1,s)$ are multiplied by
the small coupling constant $g^2_{ D^{*+}\pi^+D^0}/(16\pi)\approx
0.00289$ GeV$^2 $ which is small because $\Gamma^{tot}_{D^{*+}}
(m^2_{D^{*+}})$ is small. Numerical calculation shows that in the
$T^+_{cc}$ resonance region, the contribution of the diagrams in
Fig. \ref{Fig3} increases the width $\Gamma_{T^+_{c c}\to
D^{*+}D^0\to\pi^+D^0D^0}(s_1)$ compared to its values in the tree
approximation (see Fig. \ref{Fig2}) by about 5\%, 6\%, 2\%, and
0.6\% at $\sqrt{s_1}=3.874$ GeV, 3.875 GeV, 3.876 GeV, and 3.877
GeV, respectively. This is mainly due to the interference between
the amplitudes of the loop diagrams in Fig. \ref{Fig3} and the
amplitudes of the tree diagrams in Fig. \ref{Fig1}. Of course, the
real parts of the triangle loops play the main role in the
interference. Note that integrations in Eq. (\ref{Eq1}) with
substitutions Eqs. (\ref{Eq3}) and (\ref{Eq4}) smooth out sharp
jumps in the functions $F_{D^{*+}D^0} (s_1,s)$ and
$F_{D^{*0}D^+}(s_1,s)$. Taking into account the finite widths of the
$D^{*+ }$ and $D^{*0}$ mesons in their propagators entering into
triangle loops also smoothes the logarithmic singularities and
reduces the above estimates by approximately 10\%. As for the Schmid
theorem \cite{Sch67} about the modification of the tree contribution
by the logarithmic singular imaginary part of the loop contribution
(the modern analysis of this theorem is presented in Refs.
\cite{Guo20,DSO19}), it holds in this case to the extent that it is
permitted by the breaking of isotopic symmetry due to masses of the
particles involving into the triangle loops $D^{*+} \pi^+D^0$ and
$D^{*0}\pi^0D^+$ and also by the instability of $D^*$ mesons.

Thus, we can conclude that the tree approximation for calculating
the decay width $T^+_{cc}\to(D^{*+}D^0+D^{*0}D^+)\to\pi^+D^0D^0$,
$\pi^0D^0 D^+$, $\gamma D^0D^+$ used in Refs. \cite{Aa21a,Aa21b} is
quite sufficient at the current level of measurement accuracy.

As regards spin effects, we note the following. From general
considerations, spin effects in the essentially nonrelativistic
decay $T^+_{cc}\to D^{*+} D^0\to\pi^+D^0D^0$ should not crucially
change the assessment of the role of final state interactions. Here,
the unchanging factors are the smallness of the width $\Gamma_{D^{
*+}\to\pi^+D^0}(s=m^2_{D^{*+}})$ and positions triangle
singularities. The structure of the formulas that take into account
the loop corrections will be generally similar to Eqs. (\ref{Eq3})
and (\ref{Eq4}). There are no special factors that would enhance the
corrections (but the intermediate calculations become much more
complicated). We hope to consider spin effects somewhere else.\\

\begin{center} {\bf ACKNOWLEDGMENTS} \end{center}

The work was carried out within the framework of the state contract
of the Sobolev Institute of Mathematics, Project No. FWNF-2022-0021.

%-------------------------------------------------------------------------------------------------------------------------------------------------

\end{document}